# Hierarchical spline for time series forecasting: An application to Naval ship engine failure rate

Hyunji Moon[a], Jinwoo Choi[b]

[a]Department of Industrial Engineering, Seoul National University, mhj1667@gmail.com, 08826, 1 Gwanak-ro, Gwanak-gu, Seoul, Republic of Korea

[b]Ph.D. Candidate, Defense Management, Korea National Defense University, chlwlsdn8570@gmail.com, 33021, Hwangsanbeol-ro 1098, Nonsan, Chungcheongnam-do, Republic of Korea

**Abstract**

Predicting equipment failure is important because it could improve availability and cut down the operating budget. Previous literature has attempted to model failure rate with bathtub-formed function, Weibull distribution, Bayesian network, or AHP. But these models perform well with a sufficient amount of data and could not incorporate the two salient characteristics; imbalanced category and sharing structure. Hierarchical model has the advantage of partial pooling. The proposed model is based on Bayesian hierarchical B-spline. Time series of the failure rate of 99 Republic of Korea Naval ships are modeled hierarchically, where each layer corresponds to ship engine, engine type, and engine archetype. As a result of the analysis, the suggested model predicted the failure rate of an entire lifetime accurately in multiple situational conditions, such as prior knowledge of the engine.

## 1. Introduction

Forecasting failure rate is important as it serves as a standard for preventive measures and inventory management. Both over and underestimation of failure are detrimental to the system. Underestimation can lead to mission failure due to failures, overestimation can lead to wasted budget and reduced operational efficiency due to excessive spare part purchases. Therefore, taking account of the features of failure data into the model is important. Two characteristics of failure rate data, imbalanced category and sharing structure, are the main motivation for this paper and we propose a hierarchical spline model for improvement. First, an imbalanced category refers to the fact that the amount of data corresponding to each age or product type has a high variance. The second is sharing structure. In our case of predicting the failure rate of an engine of each ship, as engines are shared among ships, ships with the same type of engine display similar failure rate patterns. The underlying process also supports the empirical results, as the same engine types share design patterns and are made from the same factory.

Hierarchical model provides a systemic structure to address both imbalanced and sharing nature of data. In our setting, even the failure rate of an age period where data of a certain ship engine is unavailable could be forecasted by replacing its parameters with its existing correspondence. For this purpose, we have constructed the three-layer model as the following: a root layer that accounts for the core characteristics of an engine, i.e. engine archetype, a second layer which corresponds to each type of an engine, and lastly, the third layer that explains the specific characteristics of each ship.

The proposed model has additional advantages in terms of forecasting the failure of new engine types. Republic of Korea (ROK) Navy battleships evolve continuously; for example, FF (Fate Frigate) class has been replaced by FFG (Fast Frigate Guided-missile). Forecasting the failure rates of a new battleship is hard but necessary. Most existing time series models such as ARIMA or ETS (exponential smoothing) model struggles in a situation where no quantitative data exist. However, a hierarchical model can construct the outline of the failure function based on the prior qualitative information. For instance, as we will elaborate in section 5, engines constructed in a similar era show similar patterns. Therefore, information on which era the unforeseen engine was made could be utilized to forecast its failure rates.

The main contribution of this paper lies in applying hierarchical spline (HS) model to failure data from ROK Navy. Compared to the previous models, the proposed model not only improves overall forecast accuracy but also is capable of forecasting failure rates for categories with scarce data robustly. Moreover, the hypothetical similarity between each category can be tested and proved using our model; this enables users to utilize the qualitative knowledge on the unforeseen, ships with new engines for example, for forecasting. These results, when used as a reference for maintenance policy and budget allocation, could contribute greatly to the Navy's operating system. However, this model is not limited to the Naval domain. When it comes to forecasting failure rates, the circumstances where data are hierarchical, imbalanced, or insufficient are common and therefore, our model is widely applicable. For example, mechanical equipment consists of several parts. The generator, which is a part of the wind turbine, is composed of parts such as a motor and a transformer (Scheu et al., 2019) in a hierarchical structure. Using the HS model, it is also possible to predict the failure of equipment components in a hierarchical structure.

The remainder of this paper consists of five sections. Section 2 introduces the background for failure forecasting and HS model. The advantage of the chosen model is explained especially in terms of data characteristics for our setting. In section 3, details of ROK Navy data are introduced and HS model is compared with two existing models. Section 4 contains an analysis of the experimental models, and lastly, conclusions are presented in section 5.

## 2. Literature Review

In this section, imbalanced and missing nature of naval ship engine data is analyzed. Previous prediction methodologies suitable for constructing the failure function are review and the reasons for applying the Bayesian hierarchical model are described in this section. Two models are selected for comparison: ARIMA, which has been used to estimate the failure function in many past studies, and Prophet, which has been developed relatively recently but being adapted widely due to its high accuracy and scalability. Details of these methods as well as the framework for comparison are clarified.

2.1 Failure rate in Naval ship setting

Before the mission, each naval ship is equipped with a forecasted amount of spare engines. An underestimated forecast has a risk of mission failure as spares parts cannot be resupplied during mission times. An overestimated forecast may lead to reduced operating efficiency due to a load of unnecessary spare parts. Moreover, from a system point of view, overestimation induces unnecessary use of budget and even lead to inventory shortage for other ships. So, defining the optimal set of spare parts is crucial for mission success (Zammori et al., 2020).

For accurate prediction, several special features resulting from the Navy's system should be noted. First of all, imbalances are observed in two categories of the data: age period and engine types. There was only a short period of failure rate data compared to the entire lifetime. In our case, for example, an early age has less data than the rest of the age period; this might be problematic as the failure rate of young ships is needed for operation. Also, the distribution of ships for each engine type category is not balanced. In our dataset with 99 ships, there are 6, 27, 43, 19, 4 ships for each engine type category. In this case, while a satisfactory model could be obtained from an engine type with a large amount of data, other models might suffer a lack of data problems.

Moreover, the similarity between ships and engines should also be noted as they undergo the same maintenance process; planned maintenance is performed by ROK Navy regardless of the engine type (Yoo et al., 2019). Based on these circumstances, where ships as well as engines share certain qualities, the model with layered parameter structure is needed; it should be able to learn the specific structure between and within each layer from the data.

2.2 Failure forecasting models

Several models exist such as ARIMA, exponential smoothing, and seasonal trend decomposition using

Loess (Hyndman and Athanasopoulos, 2018) that could model time series characteristics of failure rate. Among the existing time series models, Prophet, which adopts Bayesian generalized additive model (GAM) shows high accuracy. Moreover, it decomposes time series into trend, seasonal, other regressor factors that enhance both its application and interpretability (Taylor and Letham, 2018). Along with the hierarchical model framework (section 2.3.), GAM is known for its ease of information sharing; when two concepts are combined by adding a hyperparameter to GAM, it becomes a hierarchical GAM (Wood, 2017; Smith, 2020) which has been applied in much research (Pedersen et al. 2019).

More specific models concentrating on the characteristics of failure have been suggested. A bathtub is a typical shape pattern observed in the failure rate. Also, Weibull or Poisson distribution are often used as a distribution of failure rate. Wang and Yin (2019) performed failure rate forecasting with the stochastic ARIMA model and Weibull distribution. Time series data have been decomposed into bathtub-shape assumed trend and stochastic factors. Parameters of the Weibull distribution were separately learned for the increase, decrease, and flat period of the bathtub. The stochastic element was obtained using ARIMA, and the time series failure rate was calculated as the sum of the trend and stochastic elements. Sherbrooke (2006) proposed Pareto-optimal algorithms, named constructive algorithms, based on Poisson distribution. However, it had limits in determining the parameter. Zammori et al. (2020) tried to solve the problem of parameter estimation of Sherbrooke's (2006) model by applying time-series Weibull distribution. Other attempts such as Pareto-optimal, Monte-Carlo (Sherbrooke, 2006), ARMA, and least-squares logarithm (Wang and Yin, 2019) have been made to add the effect of stochastic factors to this distribution.

Attempts have been made to integrate time series models with information about system architecture. In the risk analysis of deepwater drilling riser fracture (Chang et al., 2019), Bayesian network was used to predict the fracture failure rate. Bayesian network could also be used to analyze and prevent the cause of a ship's potential accidents (Afenyo et al., 2017). Time series forecasting based on Bayesian network (Dikis and Lazakis, 2019) and Analytic Hierarchy Process (AHP) (Yoo et al., 2019) illustrate these approaches. They are based on the assumption that equipment, engines for example, within the same group follow similar failure patterns.

2.3. Hierarchical model

The hierarchical model has an edge in representing the features of Navy data introduced in 2.1; imbalanced category and sharing structure, by information pooling. Gelman et al. (2005) explained that hierarchical models are highly predictive because of pooling (Gelman et al., 2013). When a hierarchical model is used, there is almost always an improvement, but to different degrees that depends on the heterogeneity of the observed data (Gelman, 2006a). When updating the model parameters, such as prior parameters, the relationship between the part of the data being used and the whole population should always be considered. Pooled effects between subclusters are partial as they are implemented through shared hyperparameters, not parameters.

By properly setting the hyperprior structure, we can find a reasonable balance between over-fitting and under-fitting, as hyperpriors are known to serve as a regularizing factor. Many examples of applying hierarchical structure in cross-sectional data exist in diverse domains, such as ecology, education, business, and epidemiology (McElreath, 2020). The structure of cross-sectional data where the whole population is divided into multiple and nested subcategories provides an excellent environment for a hierarchical model. Previous literature on comparing the education effects of multiple schools has shown that incorporating the nested structure of the state, school, and class in the model had substantial improvement in terms of accuracy and interpretability (Rubin, 1981).

Januschowski et al. (2020) has classified methods in the forecasting domain into two: global and local. Global methods jointly learn parameter using all available time series while the local methods learn independently from each time series. In this sense, hierarchical model and therefore, HS model is global while the other two compared models, ARIMA and Prophet, are local. Naturally, HS could provide the framework to forecast new types of engine that has next to no quantitative information, but as will be illustrated further, need extra care on the degree of regularization, especially for the subcategory with small amount of data. The concept of pooling could be understood in the context of the global model and is not restricted to hierarchical model; Trapero et al. (2015) achieved pooling by replacing a regression coefficient of stock-keeping units with a limited amount of data with the coefficient calculated from multiple SKUs. Other examples include recurrent neural network models with globally calculated weights (Hewamalage et al., 2020). Models that balance global and local information in the context of pooling have also been suggested, an example being pooling within each cluster (Moon and Song, 2019; Bandara et al., 2020) and two-fold spatial attention mechanism in recurrent neural network (Hewamalage et al., 2020).

2.4. Model evaluation measures

Time series cross-validation and k-fold cross-validation, along with the expanding forecast method, can be used to measure forecast accuracy in time series (Hyndman and Athanasopoulos, 2018). Several sets of training and test data are created in a walk-forward mode, and forecast accuracy is computed by averaging over the test sets. Various measures of forecast error exist, including the mean absolute, root mean squared, and mean absolute percentage error. To compare the results on different datasets, scale-independent errors including SMAPE, MAPE are preferred (Hyndman and Koehler, 2006). However, the presence of the predicted or real data in the denominator makes the measure unstable when the values take near-zero values (Hyndman and Athanasopoulos, 2018). Also, based on the case where SMAPE takes negative values, Hyndman and Koehler (2006) recommended not to use SMAPE. Based on this recommendation and as our comparison experiments are based on one set of data, we chose RMSE as our measure.

Specific to Bayesian models, measures that could diagnose the fit of a model are provided in Stan, a

Bayesian computation software. Energy Bayesian fraction of missing information (E-BMFI) and effective sample size (n_eff) are two examples. E-BMFI quantifies the efficacy of the momentum resampling between Hamiltonian trajectories and n_eff quantifies the accuracy of the Markov chain Monte Carlo estimator of a given function (Betancourt, 2017). Garby et al. (2018) shows graphical summaries based on these measures. Information criteria used to measure the fit of a model in Bayesian models include widely applicable information criterion (WAIC) and the leave-one-out cross-validation (LOOCV); they are preferred to other criteria such as Akaike information criterion (AIC) and deviance information criterion (DIC) (Vehtari and Lampinen, 2002). For Bayesian models, where the estimation of parameters is based on sampled results, it is essential to check whether chains have reached their convergence before comparing models. For these purposes, trace plots and numerical summaries such as the potential scale reduction factor, Rhat (Stan Development Team, 2017b) are used. Rhat lower than 1.1, for each parameter, is recommended. Choosing the validation set to address the sequential characteristic of time-series has also been proposed (Bürkner et al., 2020). However, as our proposed and compared models are curve fitting that does not directly address the sequential trait of time-series, except for ARIMA, we decided to use RMSE measure.

## 3. Model and Data

In this section, the aforementioned two characteristics, imbalanced and missing, are confirmed on the real dataset. Details of model construction including how basis functions and coefficients for B-spline were designed hierarchically are described.

### 3.1. Data

Data consists of 99 ship engines that are categorized into five types of engines. Therefore, our hierarchical model has a 1-5-99 structure; 1 engine archetype, 5 engine types, and 99 ship types. The numbers of ships in the five categories are also different as in section 2.1.

Fig. 1 shows the age, type of engine, and ship of existing data. As can be seen from the figure, the amount of data for each category is highly imbalanced. Moreover, the similarity between data under the same category could be inferred; for example, data with the same type of engine display a similar age period. By arranging the failure data of 99 propulsion ship engine categorized into 5 types according to their lifetimes, we got the failure rate data for the approximate total life cycle of 31 years. Note that only the records from direct maintenance workshop are included; data for warranty repair which take place at shipyard were unavailable. Due to this lowered failure count data, early period could have different pattern with the other period which could be an obstacle to partial pooling (section 4.2.1).

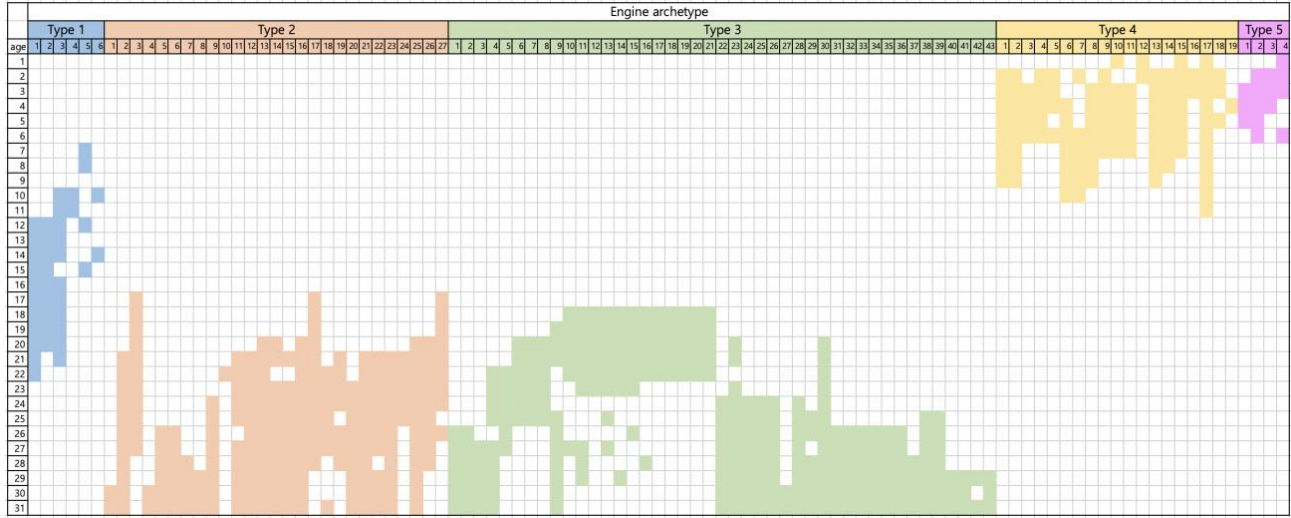

Fig. 1 Existing data by age and engine type

### 3.2. Model and process

Naval ship engines applied in the proposed model are classified as Fig. 2. The ship engine (layer 3) of each ship belongs to the same engine type (layer 2), and 5 types belong to the entire engine archetype (layer 1).

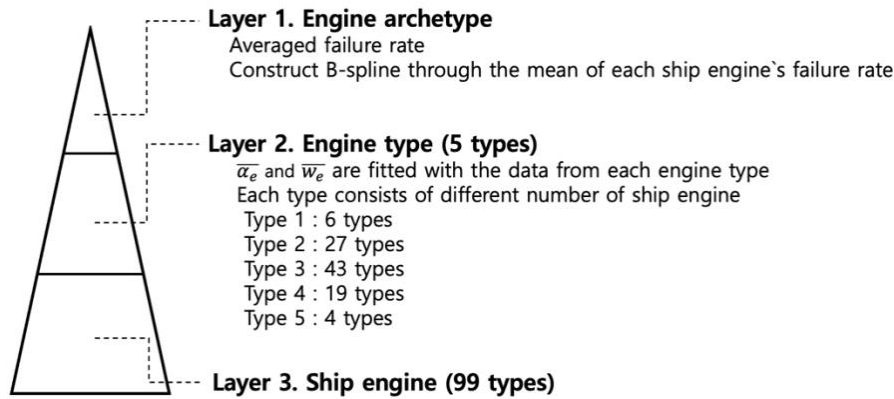

Fig. 2 Hierarchical structure of ship engine failure

Since the naval ship data is nonlinear time series data, polynomial and spline regression are considered. In polynomial regression, to achieve flexibility, a degree should be increased; however, the risk of overfitting becomes higher with its degree. To prevent this and to endow the model a form of locality, the B-Spline model is suggested: the overall life cycle, or age period, is first divided into several sections. Then low-dimensional polynomial is fitted for each section to form a piecewise polynomial spline.

The third layer of ROK naval ship hierarchy, representing the ship engine, is modeled with B-Spline. As can be seen from Equation 1, parameters for each layer share hyperparameter which leads to pooling possible. To be more specific, B-spline is pre-fitted to the averaged time series of existing data to obtain the hyperparameters, $\overline{\alpha_0}$, $\overline{w_0}$. Prior $Normal(I, 1)$ and $Normal(0, 1)$ are used and the resulting posterior means for the two hyperparameters are plugged into Equation 1. Note that, *I*, the prior mean of $\overline{\alpha_0}$ is the intercept from linear regression fit. Next, with these layer1 parameters, engine-specific parameters, $\overline{\alpha_e}$ and $\overline{w_e}$ are learned. The distribution and hyperparameter values regarding standard deviation parameters, $\sigma_w$, $\sigma_\alpha, \sigma_{\overline{\alpha}}, \sigma_{\overline{w}}, \sigma_y$ are calibrated with prior predictive checks. Also, we chose power transformation (Yeo-Johnson) to scale our data to match our prior distributions. Note that weight, w, is a vector whose length is determined by the number of knots.

$$Y_s \sim Normal(\mu_s, \sigma_y)$$
$$\mu_s = \alpha_s + \Sigma_{k=1}^{K} w_{k,s} B_k$$
$$\alpha_s \sim Normal(\overline{\alpha_e}, \sigma_\alpha)$$
$$w_s \sim Normal(\overline{w_e}, \sigma_w)$$
$$\overline{\alpha_e} \sim Normal(\overline{\alpha_0}, \sigma_{\overline{\alpha}})$$
$$\overline{w_e} \sim Normal(\overline{w_0}, \sigma_{\overline{w}})$$
$$\sigma_\alpha \sim Gamma(10, 10)$$
$$\sigma_w \sim Gamma(10, 10)$$
$$\sigma_{\overline{\alpha}} \sim Exponential(1)$$
$$\sigma_{\overline{w}} \sim Exponential(1)$$
$$\sigma_y \sim Exponential(1)$$

Equation 1

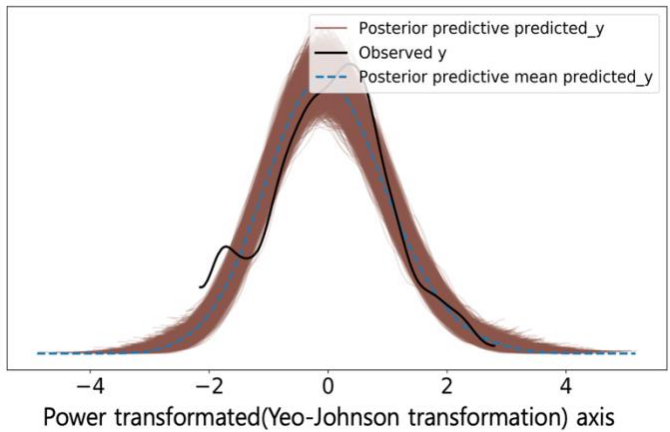

Fig. 3 Posterior predictive check

Posterior predictive check was used to validate the model. Posterior predictive check simulates data from a fitted model and compares it with the real data (Gelman and Hill, 2006). Systematic discrepancies between real and simulated data (Gelman et al., 2013) could be detected with this test. The result of our model is shown in Fig. 3.

We used Stan as a probabilistic programming language which has the advantage of fast computation owing to its efficient sampling algorithm (Carpenter et al., 2017). Code for the model is included in Appendix.

Workflow is organized as shown in Fig. 4. From failure rate data, a rough trend of the failure rate over a lifetime is deduced by averaging existing ship engine failure data from 99 ships. When ARIMA and Prophet make predictions for Layer 1, this averaged time series is used. For HS model, this trend is used to determine the value or distribution of hyperparameters. Note that the number of B-spline knot is also a hyperparameter though it does not have distribution. Apply the estimated B-spline and hyperprior to the model, and parameter fit by learning the data of each layer through MCMC sampling.

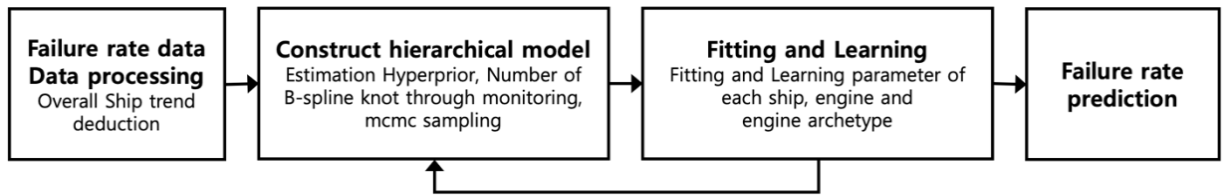

Fig. 4 Workflow

## 4. Results and Discussion

The forecast of the model are plotted and the accuracy is compared with the other two models. Accuracy comparisons are conducted in four categories, depending on application context of failure function in ROK Navy. The first two categories are in-sample tests and the last two are out-of-sample tests. We first compare the overall RMSE of 99 ship engines and secondly of each subcategory: engine type. The purpose is to observe the effect of pooling. Out-of-sample tests are divided according to whether the engine types of test data are included in the trainset or not. Different results between the model and contexts are analyzed. In addition, the similarity between the failure functions of each engine type are identified along with its implications.

### 4.1. Accuracy comparison

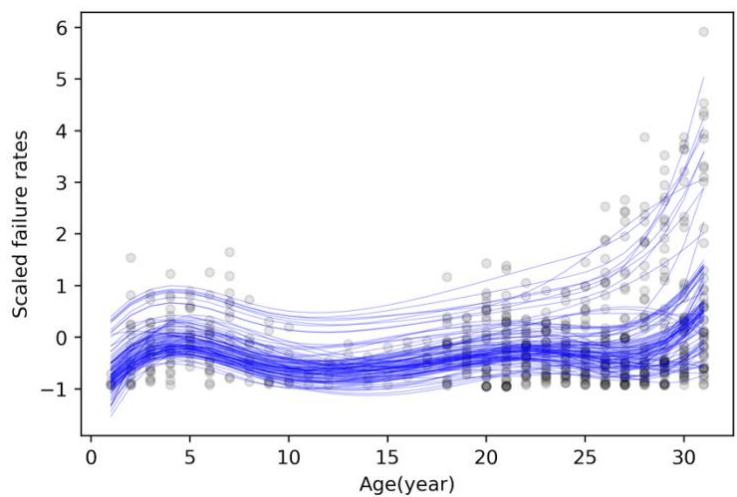

Fig. 5 Prediction results of HS model (99 ship engines)

Fig. 5 is the result of HS model predicting the total lifetime failure rate of 99 ship engines. Spots are points of failure with 99 data. Considering the missing data in the early period due to warranty repair (as in section 3.1), they are bathtub-formed, as some previous literature has suggested (as in section 2.2). Among the models introduced in section 2.2, Prophet and ARIMA are selected as baseline models each for its high accuracy and popularity, respectively. We would like to note that when auto_arima returned 0 value even though the model was not fully fit, due to lack of data for example, we have adjusted ARIMA model to exclude the periodicity and to use only moving average instead.

Accuracy comparison was performed in two ways considering the model's application situation. In general, when it is necessary to introduce a new ship engine or predict the ship engine in use, refer to the data of the same engine type. The prediction accuracy of the ship engine is used as a reference to predict the spare parts of the ship engine in use, and the prediction accuracy between the engine type and the ship engine is used as a reference when introducing a new ship engine. Therefore, first (Table 1), the accuracy of the ship engine predicted and ship engine actual values were compared, and second (Table 2), the accuracy of engine type predicted and ship engine actual values were compared. RMSE (Root Mean Square Error) was used for the error measure.

Table 1. RMSE (Ship engine predicted VS Ship engine actual)

| HS | Prophet | ARIMA |
| --- | --- | --- |
| 0.9475 | 1.0395 | 0.9553 |

Table 2. RMSE (Engine type predicted VS Ship engine actual)

| Data (number of data) | HS | Prophet | ARIMA |
| --- | --- | --- | --- |
| Type 1 (6) | 0.9761 | 1.0188 | 0.9773 |
| Type 2 (27) | 0.9567 | 0.9950 | 0.9709 |
| Type 3 (43) | 0.9697 | 0.9981 | 0.9883 |
| Type 4 (19) | 0.9426 | 1.0103 | 0.9432 |
| Type 5 (4) | 0.9593 | 1.0123 | 0.9802 |
| mean | 0.9609 | 1.0123 | 0.9720 |

The first was to compare the average by obtaining the prediction accuracy of each of the 99 ship engines. To model the 3-layer dataset, only one option exists for the hierarchical model. This is because the hierarchical model predicts 3-layer data using information of all layers. For Prophet and ARIMA, which are unable to represent the hierarchical structure, input data should be preprocessed, by averaging (as in section 3.2), to learn the parameters. As can be seen from Table 1, 2 RMSE of HS model was the lowest. RMSE in Table 1 and 2 may seem insubstantial but considering the fact that RMSE is a scale-dependent measure, it implies large difference. As mentioned previously, due to security concerns, unscaled figures are not explicitly presented. However, based on the current budget amount, at least several million dollars of budget saving are expected upon application of this model. The effect would be even greater if this model is applied to other armed force departments, Airforce and Army, which have similar structure with Navy.

## 4.2. Forecasting a new type of ship or engine

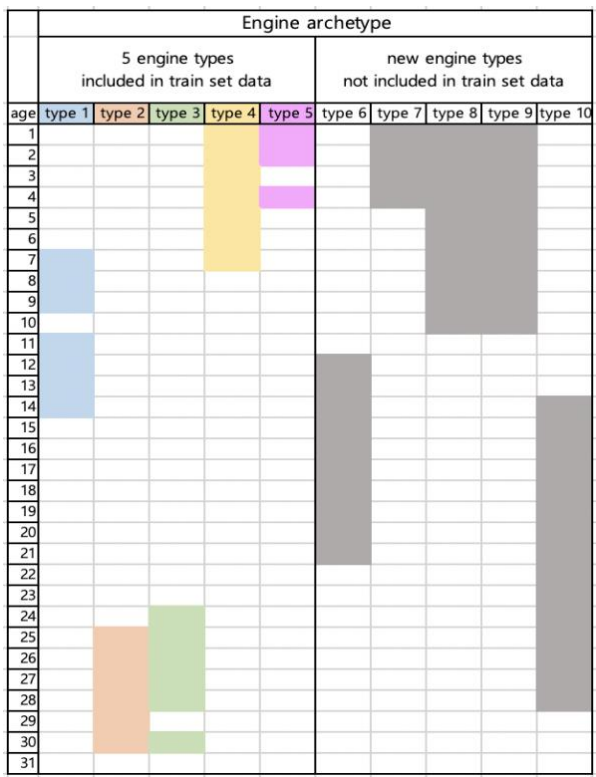

Fig. 6 Test set data

When we fit the hierarchical model with failure rates of 99 ships, the learned results are stored in the model in the form of each parameter's distribution, i.e. posterior. For example, whose prior had exponential form would evolve into a posterior distribution. Bayes formula explains this mechanism. As discussed in the introduction, engine failure rate of a new type of engine or ship is frequently needed. Depending on its engine type, the way by which the hierarchical model should be applied differs. If its engine type is present among the data, the posterior of parameters corresponding to layer 2 could be used for the forecast (4.2.1). On the other hand, if the engine type is new as well, the only information we could borrow from the previous data are posteriors of layer 1 parameters (4.2.2).

Test set data is shown in Fig. 6. Type 1 to 5 are the same as the 5 engine types included in train set data. One engine data was obtained for each engine type and prepared as a test set. Type 6 to 10 are new engine types not included in train set data. Ship engine data corresponding to 5 new engine types were prepared as a test set for each type.

### 4.2.1. New ship with its engine type included in train set

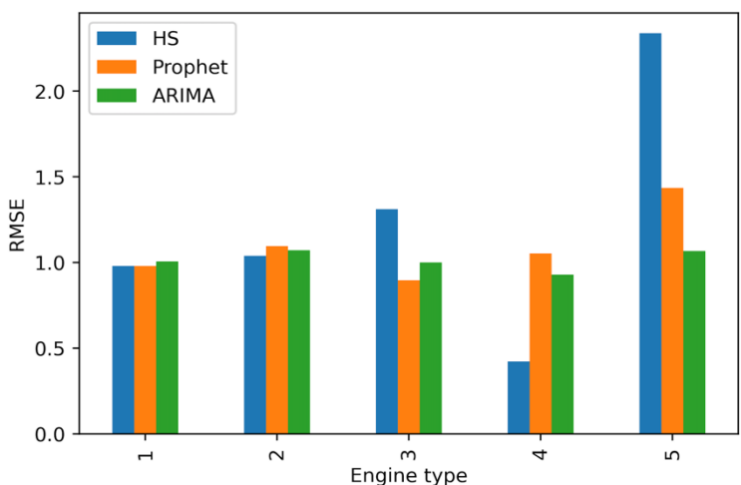

Fig. 7 RMSE of Type 1 to 5 test set

Posterior of $\overline{\alpha_e}$ and $\overline{w_e}$ could be directly used to predict engine failure of a new ship engine whose engine type is among the five trained engine types. As in section 4.1, Prophet and Arima were used as comparative models. The results are shown in Fig 7. RMSE of type 3 and type 5 for HS were higher than the other models. As mentioned in section 3.1, early period data show different pattern compared to the rest of the period

(exclusion of warranty repair). Therefore, pooling might have made the prediction less accurate by trying to shrink the prediction toward a different pattern (population mean). The fact that type 5 is a relatively minor category also contributes to this analysis; type 4 also corresponds to early period, but as it has larger amount of data (five times larger than type 5) the advantageous and disadvantageous effects of pooling could have been offset.

On the other hand, type 3 corresponds to the last age. As shown in Fig. 5, failure at the last age has high variation. Accumulated differences of usage environment could be the cause; some operator operating roughly and others stably for example. Due to this great variance, we believed test samples which only includes six instances for type 3 were not representative enough and attempted cross-validation. Fig. 8 is the summary of the process and Table 3 shows the results; it can be seen that the RMSE of the HS model is the lowest for the trained set.

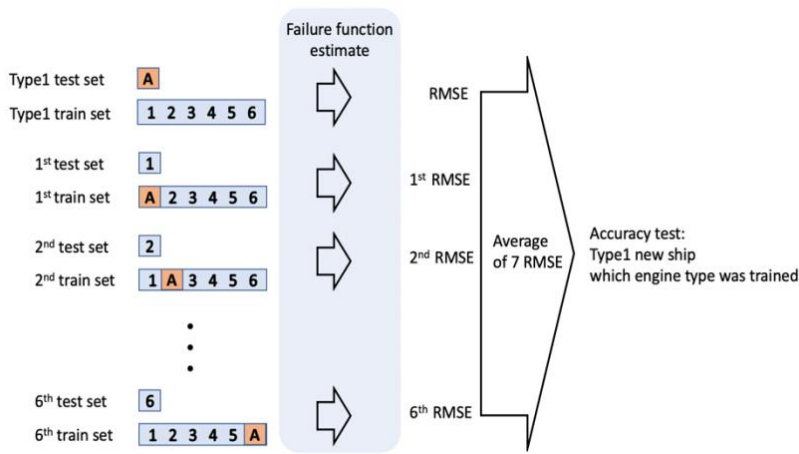

Fig. 8 Cross-Validation example of type 1

Table 3. Test set data RMSE (New ship with its engine type included in train set).

| Engine type | HS | Prophet | ARIMA |
|---|---|---|---|
| Type 1 (7) | 0.9374 | 0.9622 | 1.1886 |
| Type 2 (28) | 0.9884 | 0.9887 | 1.1151 |
| Type 3 (44) | 0.9731 | 0.9996 | 1.0471 |
| Type 4 (20) | 0.9658 | 1.0594 | 1.0342 |
| Type 5 (5) | 1.0131 | 1.2092 | 1.0962 |
| mean | 0.9756 | 1.0438 | 1.0962 |

Since the test set was added, the number of data per type was added by 1. HS model had the lowest mean

RMSE. In Section 4.1, HS model showed lower accuracy than Prophet and Arima too. Compared to section 4.1, the increase in RMSE means of engine types were the smallest in HS. The effect of hierarchical information pooling of HS model was significant when applying new data that was not learned. HS model showed lower RMSE than Arima in all types.

4.2.2. New ship with its engine type not included in train set

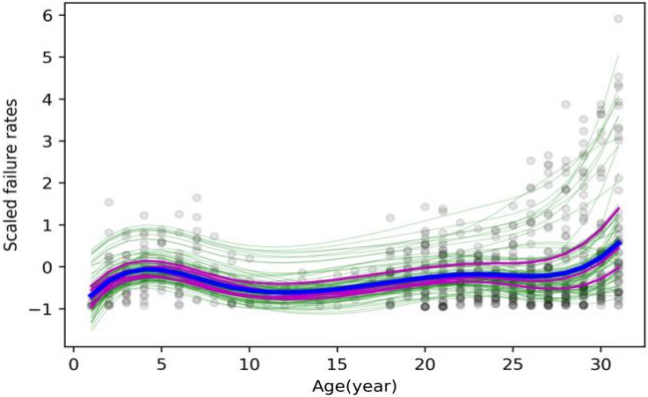

Fig. 9 HS

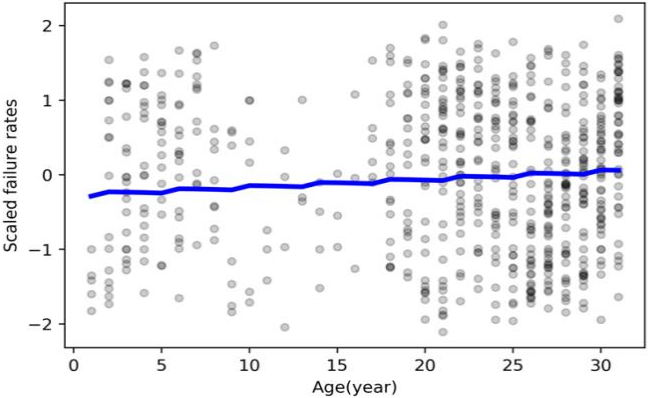

Fig. 10 Prophet

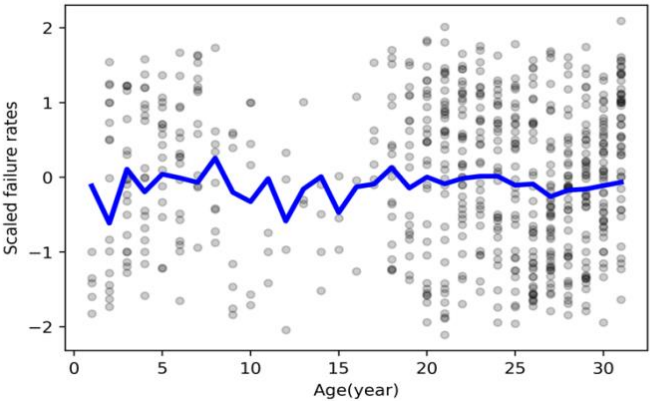

Fig. 11 ARIMA

Using HS, estimation for ship with unforeseen engine type could be performed in a robust manner; information on engine archetype is stored in hyperparameters of layer1 with which forecast can be made. In other words, the resulting $\overline{\alpha_0}$ and $\overline{w_0}$ value of a prefit, B-spline fit on averaged failure rate, are used for $\overline{\alpha_s}$ and $\overline{w_s}$ from equation 1.

Fig. 9, 10, and 11 show the prediction results of HS and comparative models. Dotted point are train data. From Fig. 9, green, red, and blue line each correspond to the predicted failure of each ship engine, engine type, and engine archetype from HS model. Blue line from Fig. 10 and 11 is Prophet and ARIMA's prediction of the entire lifespan.

For the engine type with no historical data, failure function of similar types (layer 2) or engine archetype (layer 1) could provide valuable information for its prediction. These situation are frequent since technology develops and there is a constant need to replace or upgrade the engine. We confirmed that HS performed well when only the information on engine archetype are available; in other words even the type of the engine was not included in the train set (Table 4). This is notable because it is the most difficult, but necessary, case in real situation.

Table 4. Test set data RMSE (New ship with its engine type not included in train set).

| Model type | HS | Prophet | ARIMA |
|---|---|---|---|
| Type 6 | 0.9620 | 1.0155 | 1.1078 |
| Type 7 | 0.7782 | 1.0629 | 0.7961 |
| Type 8 | 1.1069 | 1.0563 | 1.0263 |
| Type 9 | 0.8546 | 1.0352 | 1.0031 |
| Type 10 | 0.9128 | 1.0064 | 0.9991 |
| Mean | 0.9229 | 1.0353 | 0.9865 |

In all new types, Except for type 8, the HS model had a lower RMSE than the comparative model. Type 8 can be the effect of initial data as in Section 4.2.1. Types 7, 8 and 9 are all initial data, and their RMSE rankings are different. In Table 4, HS has the lowest average RMSE. Using a new type of ship engine is always a concern for the Navy. This is because the budget for maintenance cannot be estimated. It can be helpful in this situation that the mean of the HS model is the lowest in Table 4. This would not be a problem only for the ROK Navy.

In test set data (Fig. 6), data of type 1, 2, 3, 6, and 10 are commonly ship engines with an age of 5 years or more. These types showed the lowest RMSE of test set prediction results (Table 3, Table 4). In Section 3.1, the data of the initial part is said to reflect fewer data than the actual (because the data in this study only include the military direct maintenance workshop). In other words, it can be said that the data of the initial part is less reliable than other sections of data. In general, the warranty repair period of the ROK naval ship engine does not exceed 5 years. Type 1, 2, 3, 6, and 10 do not include data for the initial 5-year period with relatively low reliability. That is, the HS model proved better performance than the comparative model by showing a low RMSE in the test set data of all engine types with relatively reliable data. As shown type3 in Fig. 7, there is a case where the accuracy is slightly degraded due to the increased deviance at the end of the life. However, through the results of Table 3, it was confirmed that the RMSE of the HS model was low at the end of its life. In addition, the difference in deviance at the end of life is estimated to be less than the effect of low reliability at the beginning of life.

4.3. Reflecting the qualitative knowledge

Prediction can be improved in the presence of the qualitative knowledge, construction era of the new engine type, for example. This act of translating qualitative into quantitative knowledge could be justified by analyzing their relationship with the existing failure functions of five engine types. Fig. 12 and Table 5 give interpretable results. Engine failure function is largely influenced by the construction era. Based on the historical data, we have classified Type 4 and 5 as Early, Type 1 as Middle, and Type 2 and 3 as Last. As shown in Table 5, in general, the Euclidean distance between Early and Middle is small compared to Early and Last. This could be understood in terms of continual development of engine technology and supports the result of our model. To be more specific, similar construction era resulted in a similar trend in failure function with the exception of Type2 and Type3. The distance between these old engines is large. We think this could be the result of accumulated differences in ship usage environment. Early aged engines would show similar failure patterns between types compared to older engines. Other factors including environmental (East or West sea) and purpose (shipping or guarding in the frontal line) factors would affect failures in old aged engines greatly. Second, is the technology development. It can be said that the latest engines have similar failure functions. Type 4 and 5 engines were constructed after 2010 while type 2 and 3 engines were constructed in the 1990s.

Based on the qualitative knowledge on the closeness of new engine type with the existing engine types, posterior of $\overline{\alpha_e}$ and $\overline{w_e}$ of previous engine types could be used as a hyperprior for new ship's $\overline{\alpha_s}$ and $\overline{w_s}$. Compared to using the original prior $Normal(\overline{\alpha_0}, \sigma_{\overline{\alpha}})$ and $Normal(\overline{w_0}, \sigma_{\overline{w}})$ from equation 1, this would give more accurate results as more prior knowledge could be reflected for the prediction.

Table 5. Euclidean distances

| Engine type | | Euclidean distance |
|---|---|---|
| Early vs Middle | 1 vs 5 | 0.0425 |
| Early vs Early | 4 vs 5 | 0.0461 |
| Early vs Middle | 1 vs 4 | 0.0466 |
| Middle vs Last | 1 vs 3 | 0.1057 |
| Early vs Last | 2 vs 5 | 0.1076 |
| Middle vs Last | 1 vs 2 | 0.1090 |
| Early vs Last | 3 vs 5 | 0.1237 |
| Early vs Last | 2 vs 4 | 0.1244 |
| Early vs Last | 3 vs 4 | 0.1278 |
| Last vs Last | 2 vs 3 | 0.1836 |

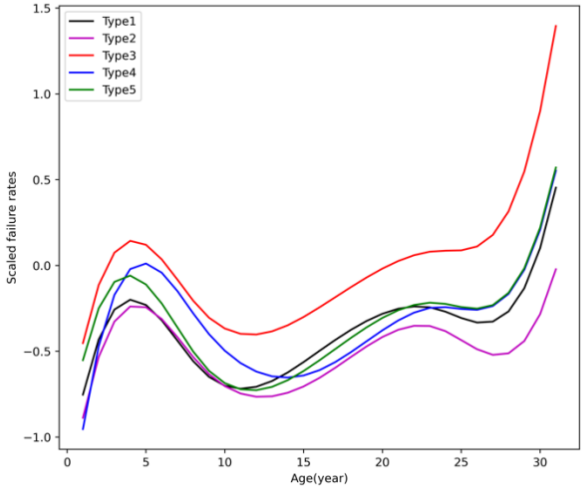

Fig. 12 Distance between engine types

## 5. Conclusions

We have proposed using HS to develop a hierarchical model for forecasting failure rates. This approach shines especially when the data are imbalanced and hierarchically structured. We demonstrated the applicability of the model using a real-world dataset of failure rate data from Naval ships and compared it with previous methods. Through these comparisons, we confirmed that the prediction performance of our novel model in the given dataset was greatly improved. Moreover, we have shown how qualitative knowledge, such as the belonging to the same series or construction era, could be incorporated into the model; this approach was justified by further analyzing the relationship between each parameter. These techniques could greatly improve Naval ship management efficiency.

Some improvement could be noted for further studies. First, prevention repair which may affect the failure pattern could be considered. A more advanced model that incorporates the probability of failure after the prevention repair is needed to design a model. Second is convergence and evaluation measures. There were few instances with low E-BFMI and effective sample size, n_eff. Improving the model in terms of higher E-BFMI and n_eff measures would result in a better fit of the model. Thirdly, due to substantial operational differences between combat and non-combat ships, only combat ships are used in this paper. However, if the differences could be incorporated in the further models, by using categorical variables, a more accurate model could be possible based on a larger amount of data.

HS can contribute greatly to the following areas. First, failure rate prediction could be used as a quantitative reference when establishing a maintenance policy. Proper maintenance not only improves the availability and mission completion rates but also reduces the budget by reducing unnecessary maintenance. Second, from a broader perspective, the predicted failure trend can be a qualitative reference for designing the optimal life cycle of a ship. For instance, based on our results, the failure rate increases dramatically as the ship becomes senile. Therefore, optimal retirement period could be decided by balancing the maintenance and construction costs.

**Appendix: Stan code HS model**

```
data {
  int <lower = 1> K;
  int <lower = 1> N;
  int <lower = 1> T;
  int <lower = 1> S;
  int <lower = 1> E;
  int <lower = 1> Age[N];
  int <lower = 1> Ship[N];
  int <lower = 1> S2E[S];
  matrix[T,K] B;
  real mu_a_bar;
  real mu_w_bar[K];
  vector [N] Y;
```

```
}

parameters {
   vector[S] a;
   real a_bar[E];
   vector[K] w[S];
   vector[K] w_bar[E];
   real<lower=0> s_a;
   real<lower=0> s_w;
   real<lower=0> s_a_bar;
   real<lower=0> s_w_bar;
   real<lower=0> s_Y;
}

transformed parameters {
   vector [N] mu;
   for (n in 1: N){
      mu[n] = a[Ship[n]] + B[Age[n]] * w[Ship[n]];
   }
}

model {
   s_a ~ gamma(10,10);
   s_w ~ gamma(10,10);
   s_a_bar ~ exponential(1);
   s_w_bar ~ exponential(1);
   s_Y ~ exponential(1);

   for (s in 1:S){
      a[s] ~ normal(a_bar[S2E[s]], s_a);
      w[s] ~ normal(w_bar[S2E[s]], s_w);
   }

   for (e in 1:E){
      a_bar[e] ~ normal(mu_a_bar, s_a_bar);
```

```
      w_bar[e] ~ normal(mu_w_bar,s_w_bar);
   }

   Y ~ normal(mu, s_Y);
}
generated quantities{
   vector[N] log_likelihood;
   for (i in 1:N) {
      log_likelihood[i] = normal_lpdf(Y[i]|mu[i], s_Y);
   }
}
```